\documentclass[12pt,preprint]{aastex}

\begin{document}

\title{Late Light Curves of Normal Type Ia Supernovae}

\author{Jessica C. Lair}
\affil{Department of Physics and Astronomy, Eastern Kentucky University, Richmond, KY 40475}

\author{Mark D. Leising}
\affil{Department of Physics and Astronomy, Clemson University, Clemson, SC 29634}

\author{Peter A. Milne and G. Grant Williams}
\affil{Steward Observatory, University of Arizona, Tucson, AZ 85721 }

\begin{abstract}

We present late-epoch optical photometry (BVRI) of seven normal/super-luminous Type Ia supernovae: SN 2000E, SN 2000ce, SN 2000cx, SN 2001C, SN 2001V, SN 2001bg, SN 2001dp.  The photometry of these objects was obtained using a template subtraction method to eliminate galaxy light contamination during aperture photometry.  We show the optical light curves of these supernovae out to epochs of up to $\sim 640$ days after the explosion of the supernova. We show a linear decline in these data during the epoch of 200-500 days after explosion with the decline rate in the B,V, \& R bands equal to about 1.4 mag/100 days, but the decline rate of the I-band is much shallower at 0.94 mag/100 days.  

%The data are compared to model curves with two different methods of Type Ia supernova modeling, a radiation transport model with complete trapping of the positrons and a energy deposition model including positron transport, with neither of the models fitting the light curves well, particularly the I-band curve.      

\end{abstract}

\keywords{supernovae : general}

\section{Introduction}

Type Ia Supernovae (SNe Ia) are thought to be the thermonuclear explosions of white dwarf stars \citep[see][and references therein]{2000A&ARv..10..179L}.  The light curves of SNe Ia are powered by the deposition of energy into the SN ejecta from the gamma-ray and positron products of the $^{56}Ni$ $\rightarrow$ $ ^{56}$Co $ \rightarrow $ $^{56}$Fe decay \citep*{1969ApJ...157..623C}. The extreme brightness and seemingly uniform light curves of SNe Ia make them good candidates for use as standard candle distance indicators, as first suggested by \citet{1939PhRv...55..726Z}. In more recent years, it has been shown that Type Ia supernovae do not have entirely uniform peak luminosities, light curve shapes, or spectra.  They can, however, be normalized to account for inhomogeneity, thus allowing these objects to be used as standard candle distance indicators \citep[e.g.][]{1993ApJ...413L.105P,1996ApJ...473...88R}.  Light curves can be separated into three luminosity categories, normally luminous (N), super-luminous (SP), and subluminous (sb), based on light curve shape, which also correlates with some spectral features (see \citet{2001ApJ...546..734L} for a study of the frequency of each sub-class).  One measure of these classes is the \citet{1993ApJ...413L.105P} luminosity-width relation, $\Delta m_{15}(B)$, which is defined as the decline in B-band magnitude in the first fifteen days after peak, where slower declining SNe are brighter than faster declining SNe.  An empirical relationship for the correlation between $\Delta m_{15}(B)$ value and late light curve shape was derived by \citet*[hereafter MTL01]{2001ApJ...559.1019M}, showing the break between the N/SP and the sb classes to be 1.60.  In that study there were no SNe in the data set with $\Delta m_{15}(B)$ values between 1.50-1.68 so the exact location of the break is undetermined.  The issue is further complicated by SN 1986G, which has $\Delta m_{15}(B)=1.73$, but shows characteristics of sb SNe up to 200 days after explosion but N/SP SNe after 200 days \citep{2005coex.conf..183M}.  For this work, we consider all SNe Ia which have  $\Delta m_{15}(B)$ values smaller than 1.68 to be normally- or super-luminous.  The sub-classes of SNe Ia were also spectroscopically defined by \citet{1993AJ....106.2383B}.

Between 100-200 days after the explosion the ejecta become transparent to the $^{56}$Co $\rightarrow$ $^{56}$Fe-decay  $\gamma$-rays and the light curve is powered by the deposition of the positron kinetic energy into the ejecta. This work concentrates solely upon this ``positron phase" during which the energy deposition is dominated by the slowing of the energetic positrons via collisions with the bound and free electrons of the ejecta. The escape of a fraction of these positrons from the ejecta has been suggested as a possible source of the Galactic 511 keV annihilation radiation provided certain conditions exists in the ejecta
to permit escape \citep*[hereafter MTL99]{1993ApJ...405..614C,1999ApJS..124..503M}.  At late epochs, gamma-rays escape the ejecta in abundance, leading to the deposition of kinetic energy from non-thermal positrons dominating the total energy deposition.  Many groups have done work in simulation and observation during this epoch in  attempts to determine if positrons do escape the ejecta in significant quantities, or alternatively are trapped to deposit 100\% of their kinetic energy and annihilate in the ejecta.  The primary goal of observing the late emission of SNe Ia is to determine whether some of the positrons do escape the ejecta, or whether all are trapped to annihilate and deposit their kinetic energy into the ejecta. These studies are based upon demonstrating that:  1) the decay products of $^{56}$Co~$\rightarrow$~$^{56}$Fe decays can acceptably account for the spectra and light curves of SNe Ia, and more subtly, that 2) the energy deposition rate of the escape or trapping simulations provides the superior agreement with the observations than the other. In this work, we present BVRI light curves of seven normally- or super-luminous SNe Ia.  The goal of these observations is to better understand late optical light curves, which will eventually lead to the correct determination of whether positrons escape the ejecta of this sub-class of SN Ia events.

\section{Positron Transport and Energy Deposition}

Positrons are created in SN Ia ejecta via the decay chain of $^{56}$Ni $\rightarrow$ $^{56}$Co $\rightarrow$ $^{56}$Fe. Whether any of these positrons escape the ejecta into the Galaxy is an unsolved problem, a problem that can be addressed through determining the energy deposition rate in the SN ejecta. As positrons are charged particles, their movement through the SN ejecta is determined by the magnetic field in the ejecta. A weak magnetic field and/or a radially-combed magnetic field would permit the $^{56}$Co $\rightarrow$ $^{56}$Fe decay positrons to diffuse through the ejecta with an increasing fraction being able to escape the ejecta after roughly 100 days, a possibility suggested by \cite{1979ApJ...230L..37A}. \citet*{1980ApJ...237L..81C}, \citet*{1993ApJ...405..614C}, \citet*{1998ApJ...500..360R}, \citet{1997A&A...328..203C}, MTL99 and MTL01 have all performed simulations assuming a weak and/or radially-combed field and support the suggestion that a small fraction of these positrons could 
escape. 

By contrast, a strong, tangled magnetic field would trap all positrons and lead to complete deposition of positron kinetic energy into the ejecta. The positrons would develop non-negligible lifetimes after $\sim$150 days, but eventually almost 100\% of the positron energy would be deposited into the ejecta. This mechanism to temporarily store decay energy in the form of positron kinetic energy makes the energy deposition rates for this trapping scenario larger at late times than the assumption of {\it in-situ} deposition of 100\% of the positron kinetic energy that is often made for SN Ia. We note that the energy deposition rates for the trapping scenario and {\it in-situ} assumption are very similar compared to the difference between these rates and the rate for the positron escape secenarios.  

Energy deposition will differ between the two scenarios in two ways. First, the total energy deposition rate will be higher for the trapping scenario, as in the escape scenario positrons increasingly escape, carrying away energy that is deposited in the trapping scenario. Second, in the escape scenario, positrons diffuse to non-$^{56}$Ni rich regions of the ejecta, whereas in the trapping scenario the positrons remain in the iron-rich regions of the ejecta. This diffusion leads to energy deposition in regions that differ in both composition and density compared to the trapping scenario.  For the trapping scenario, the only source of energy diffusion comes from the interactions of the gamma-ray photons, which at that epoch are the secondary energy deposition source.  These scenarios can thus potentially be distinguished both through comparisons of the bolometric light curves with simulated energy deposition rates, and through detailed study of the late-epoch spectra (studying both the elements that contribute to the spectra as well as their ionization states).   

\section{Radiation Transport Simulations}

Energy deposition in the ejecta is what gives rise to the emission from the SN, but detailed radiation transport simulations must be performed in order to generate the full spectrum of light that emerges from the SN ejecta. The interactions of positrons with the ejecta give rise to many ionized and excited electrons for each positron, electrons that recombine to produce the optical, UV, NIR, IR emission. The SN is constantly expanding such that no region could be considered in equilibrium, making estimation of the detailed ionization structure complex. This is made even more difficult by the imperfect knowledge of the atomic physics that govern the various transitions that give rise to the forbidden lines of cobalt and iron that dominate the late, nebular spectra. Two early efforts at simulating the late emission from SNe Ia were performed by \citet{1980PhDT.........1A} and \citet*{1980ApJ...237L..81C}. The two efforts treated the radiation transport quite differently, and will be explained separately. 

\citet{1980PhDT.........1A} evolved a model to $\sim$700 d post-explosion assuming {\it in-situ} energy deposition and that the ionization state of the ejecta is dictated by collisional processes (and thus could be characterized by a temperature). He found that one consequence of the radiation transport was that the ejecta progressively cooled with time, leading to an increasing fraction of the emission coming out in the infrared. As the infrared photons only weakly interact with the ejecta, this mechanism efficiently cools the ejecta after roughly 500 days post-explosion. Since the infrared wavelengths have historically been unobservable, this phenomenon was named the ``infrared catastrophe", or ``IRC". Comparisons between energy deposition rates and observations would require either complete coverage of the UV/OPT/IR wavelength ranges, or detailed modeling of the evolution of the bolometric correction of each photometric band with time. Stated differently, the optical bands became increasingly poorer tracers of the bolometric luminosity (and thus the energy deposition rate) with time. Included in that study was a comparison with a $\sim$700 days optical spectrum  SN 1972E, which was suitably explained by the model spectrum. The core-collapse SN 1987A, the best observed SN in history, exhibited light curves and spectra that were well-fitted with simulations that feature cooling as with the IRC. Lowering of the ionization state with time was thus observed for a core-collapse SN, which motivated efforts to simulate a SN Ia with the same code.  \citet{1996ssr..conf..211F}, authors of the highly successful fits to SN 1987A, studied the IRC and SN Ia light curves in great detail, finding that the code that reproduced the emission from SN 1987A generated a steep drop in the optical light curves beginning at roughly 500 days for SN Ia models, a feature inconsistent with the UBV photometry of SN 1972E. A number of ideas as to what might lead to the failure of the simulations were explored, 1) whether shortcomings of the atomic physics (i.e. estimates of the transition probabilities) falsely shorten the timescale of the onset of the IRC, 2) whether SNe Ia possess additional energy input sources not treated in the models (more $^{56}$Ni produced in the explosion and/or additional radioactivities), 3) asymmetries not treated in the 1D simulations (global or otherwise) produce an excess at late times. When combined with the lack of observational evidence of ejecta cooling, it was unclear whether color evolution occurred at significant levels in SNe Ia, thus the problem was left as unresolved in that work.

By contrast, \citet*{1980ApJ...237L..81C} suggested that the radiative transitions occurred on a shorter timescale than collisions during the nebular phase, and that the ionization state would not change appreciably at late times. If this were indeed the case, then photometry in any single band would trace the bolometric luminosity (and thus the energy deposition rate) accurately enough to differentiate between positron transport scenarios.  Colgate, Petschek \& Kreise fitted simulations to the B-band photometry of SN 1937C \& 1972E, and concluded that the light curves supported the positron escape scenario.  A number of subsequent works have performed either positron transport or radiation transport, comparing their simulations with observed photometry and spectra. \citet{1997A&A...328..203C} determined, using V-band light curves as tracers for the bolometric luminosity, that varying degrees of positron escape occur for different SNe Ia. \citet*{1998ApJ...500..360R} compared pseudo-bolometric light curves generated from BVRI photometry for SN 1972E, SN 1991bg, and 1992A to models and saw suggestions of positron escape in some SNe Ia. MTL99 studied the V and B band light curves of 10 SNe Ia and concluded that for the normally- and super-luminous SNe Ia, positron escape better explained the light curves than positron trapping. That study was followed by a study of the BVRI light curves of 22 SNe Ia (MTL01). The second study confirmed the differences between the late light curves of normally- and super-luminous versus subluminous SNe Ia. It also showed that a pseudo-bolometric light curve created from the BVRI light curves is steeper than the trapping energy deposition rate, but roughly equal to the positron escape energy deposition rate.  All of those studies discussed discriminating between positron escape scenarios, but employed only optical light curves, in particular V and B band photometry. Implicit in all of those comparisons was the assumption that color evolution was not significant enough to invalidate the use of optical light curves to trace the bolometric light curve (and thus the energy deposition rate).  With the study of SN 2000cx, it became clear that studies of the optical wavelength range are not sufficient to characterize the bolometric light 
curves of SNe Ia.

\subsection{SN 2000cx and NIR Photometry}

A fundamental prediction of published radiation transport simulations is that the iron-peak elements will transition from doubly-ionized to singly-ionized to neutral with time. That leads to transitions from emission primarily in the optical wavelength range to an increasing fraction in the NIR, then (at still later times) to emission primarily in the IR. Thus, hints of the impending IRC could potentially be seen in a flattening of the NIR light curves. Observations of SN 2000cx between 350 - 500 days post-explosion in the optical and NIR afforded the first opportunity to search for these transitions. \cite{2004A&A...428..555S} showed photometry that for the first time included NIR emission after 300 days. Whereas the optical photometric bands are similar to other SNe Ia (suggesting that SN 2000cx was not anomalous at late times), the NIR light curves were flat, showing no fading at all during that epoch. In addition, the I band was found to have a shallower slope than the B,V \& R bands. A bolometric light curve created from that BVRIJH photometry suggested that the bolometric light curve is flatter than the V band light curve.  Thus, while the V band follows a steeper slope consistent with the slope of the energy deposition rates with positron escape, the bolometric light curve follows the shallower slope that suggests positron trapping.  

\cite{2004A&A...428..555S} then fitted radiation transport simulations of BVRIJH light curves to the SN 2000cx data.  The simulations featured optical light curves that were steeper than the NIR light curves, although the simulated light curves had a strong dependence upon the uncertainty about how to treat the photoionization.  As the model assumed instantaneous, {\it in-situ} deposition of positron kinetic energy, these simulations presented a scenario where the light curves of SNe Ia are consistent with positron trapping. The simulations presented in \cite{2004A&A...428..555S} featured an IRC delayed to later epochs than previous simulations, but by 600 days the optical and NIR bands were predicted to rapidly fade, a tendency that was not consistent with the latest epoch observations of SN  2000cx, nor with 600+ day optical observations of other SNe Ia. The version of the simulations that included photoionization featured a delayed fading of the optical bands, which is important to explain the $\sim$ 700 day V band detection of SN 2000cx, as well as optical band detections of other SNe Ia. The simulations also fitted a 360 day spectrum of SN 2000cx, arriving at the slightly contradictory finding that the version of the simulations {\bf without} photoionization provided a better fit than including photoionization. Indeed, a number of other spectral studies appear to produce better fits to nebular spectra than the fits shown in \cite{2004A&A...428..555S} \citep[e.g.][]{1997MNRAS.290..663B,1997ApJ...483L.107L}. Nonetheless, the spectral fits are certainly good enough to provide bolometric corrections at the level of accuracy to combine the BVRIJH light curves into a UVOIR bolometric light curve.  

SN 2000cx was considered an anomalous SN Ia, as the near-peak B-band photometry crossed the standard SN Ia templates \citep{2001PASP..113.1178L}. It has therefore been considered possible that features of the late light curves of SN 2000cx are not typical of SNe Ia in general. One argument against that view is that the J \& H band photometry of SN 2000cx appears to be consistent with H-band spectro-photometry of SN 1998bu \citep{2004A&A...426..547S}. NIR spectra of SN 1998bu were fitted with a model dominated by Fe II and Co II features, confirming that the NIR emission is due to singly-ionized iron-peak elements. \citet{2004A&A...426..547S} introduce the possibility of a partial cooling, where the iron-peak elements cool to the singly ionized state, but not to the neutral state (at least during the observed epochs). Two consequences of that possibility would be: 1) the optical and NIR light curves not falling steeply and 2) the IR emission would not brighten (i.e. there is color evolution into the NIR, but there is no IRC).  Another study shows that the normally-luminous SN 2003cg ( $\Delta m_{15}(B)=1.12 \pm 0.04$) also exhibits a bright NIR/optical ratio at 400 days after the peak luminosity \citep{2006astro.ph..3316E}. These observations suggest that the late NIR emission from SN 2000cx was not anomalous. As SN 2000cx was one of the SNe observed in this study, we will further show that the late optical emission from SN 2000cx is also not anomalous. First we present the observations and derived quantities, and then we discuss our findings in the context of the energy deposition and ionization state of the SN ejecta.

\section{Observations}

\begin{figure}
\plotone{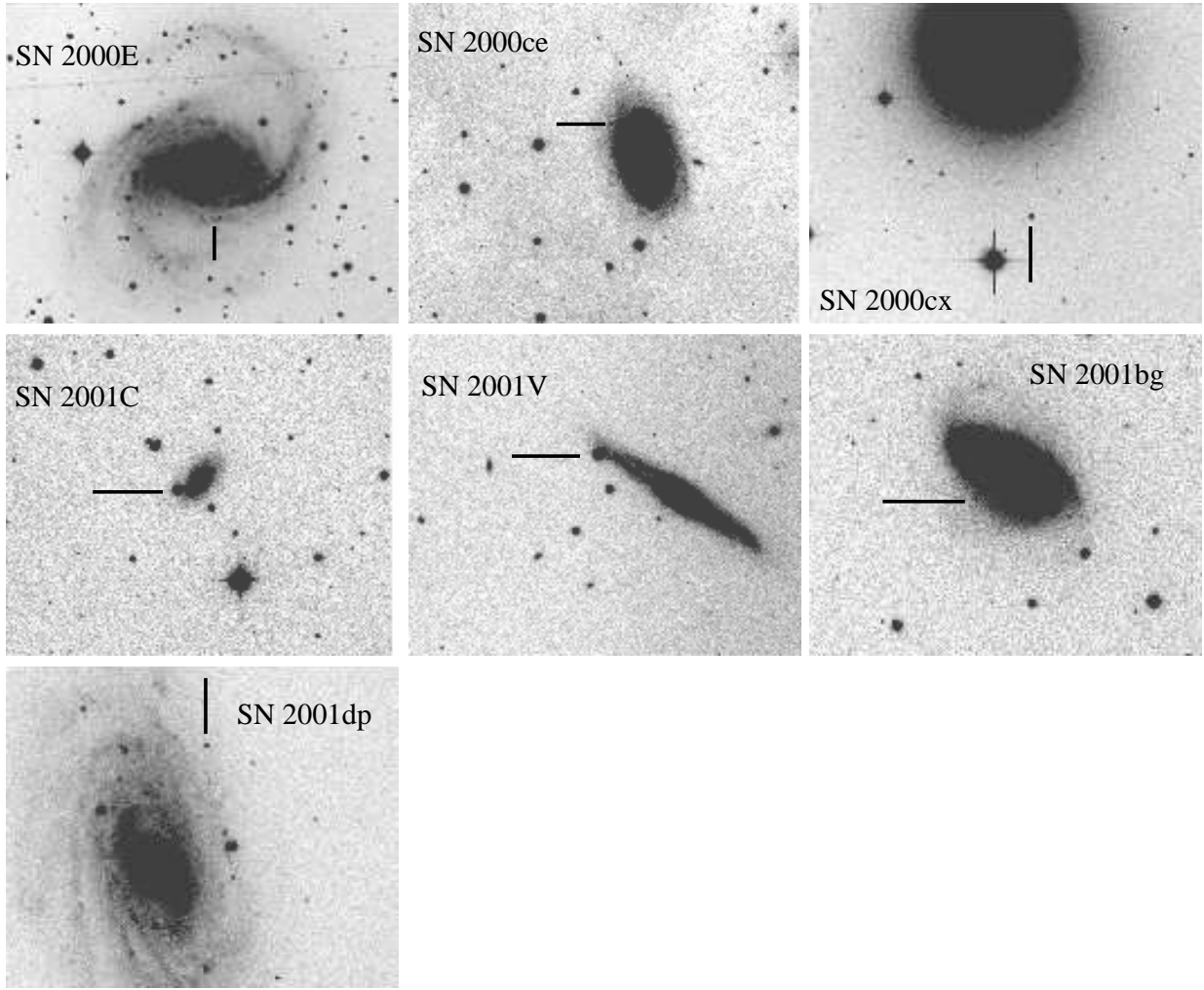}
\caption{Type Ia Supernovae in this study.  Each image has a field of view of 5'x5' with North being up and East being to the left. \label{mosaic} }
\end{figure}

The data for this project were obtained with both the Bok 90" (2.3-m) telescope on Kitt Peak and the Kuiper 61" (1.6-m) telescope on Mt. Bigelow\footnote{For more information on the detectors used see ccd21, ccd24, and ccd32 at: http://www.itl.arizona.edu/UAO/FacilityCCDs.htm}.  Figure \ref{mosaic} shows the SNe observed in this study.

\clearpage

\clearpage

SN 2000E was discovered during observations of SN 1999el in NGC 6951 on 2000 January 26.73 UT \citep{2000IAUC.7351....1V}.  The supernova was located at $\alpha = 20^h37^m13^s.8, \delta = +66^\circ05'50''.2$, which is $6''.3$ west and $26''.7$ south of the center of NGC 6951.  A spectrum obtained on 2000 January 27.83 UT showed that SN 2000E was a Type Ia supernova a few days before maximum luminosity \citep{2000IAUC.7351....1V}.  \citet{2003ApJ...595..779V} determined the $\Delta m_{15}(B)$ of SN 2000E to be $0.94 \pm 0.05$ and the reddening to be $E(B-V)= 0.50 \pm 0.09$.  

%Figure \ref{sn2000E} shows our R-band image of SN 2000E obtained on October 18, 2000 with the Bok telescope, showing the SN at 278 days after explosion.  

%\clearpage

%\begin{figure}
%\plotone{f1.ps}
%\caption{SN 2000E in NGC 6951.  R-band image obtained on Oct. 18, 2000.\label{sn2000E}}
%\end{figure}

%\clearpage

SN 2000ce in UGC 4195 was discovered on 2000 May 8.13 UT \citep{2000IAUC.7417....1P}.  The supernova was located at $\alpha = 8^h05^m09^s.45, \delta = +66^\circ47'15''.2$, which is $15.''1$ east and $17.''3$ north of the center of UGC 4195.  A spectrum taken on 2000 May 10.14 UT  showed it to be a Type Ia supernova 6 $\pm$ 3 days after maximum \citep{2000IAUC.7422....2J}.  \citet{2001AJ....122.1616K} performed early observations of SN  2000ce and determined the $\Delta m_{15}(B)$ value to be $0.99 \pm 0.10$ and the extinction to be $A_V = 1.67 \pm 0.20$.   

%Figure \ref{sn2000ce} shows our V-band image obtained November 16, 2000 with the Bok telescope, showing the SN 218 days after explosion.  

%\clearpage

%\begin{figure}
%\plotone{f2.ps}
%\caption{SN 2000ce in UGC 4195.  V-band image obtained on Nov. 16, 2000\label{sn2000ce}}
%\end{figure}

%\clearpage

SN 2000cx was discovered in NGC 524 on 2000 July 17.5 UT \citep{2000IAUC.7458....1Y}.  The supernova was located at $\alpha = 01^h24^m46^s.15, \delta = +09^\circ30'30''.9$, which is 23".0 west and 109".3 south of the nucleus of the S0 host galaxy.  A spectrum obtained on 2000 July 23 UT showed SN 2000cx to be a peculiar Type Ia supernova, which greatly resembled the overluminous SN 1991T \citep{2000IAUC.7463....1C},  showing prominent Fe III absorption lines near 4300 and 4900\AA and a weak Si II line at 6150\AA.  The early light curve of SN 2000cx was also very unusual.   \citet{2001PASP..113.1178L} showed an asymmetry in the peak of the B-band light curve.  Pre-maximum, the light curve follows almost exactly SN 1994D, which is considered a normal SN Ia, but post-maximum it follows SN 1991T.  This results in a $\Delta m_{15}(B)$ value of $0.93 \pm 0.04$ \citep{2001PASP..113.1178L}, which is very similar to the value for SN 1991T of $0.95 \pm 0.05$ \citep{1998AJ....115..234L}.  \citet{2001PASP..113.1178L} also determined that there is no host galaxy reddening of SN 2000cx due to the SN position in the galaxy and the lack of the narrow interstellar Na I D absorption line.  The total reddening is just the Galactic reddening of $E(B-V) = 0.08$ \citep{1998ApJ...500..525S}.  \citet{2004ApJ...606..413B} suggest that the asymmetry of the B-band peak could be explained, in part, by high-velocity features in the spectrum.  They suggest these features could be caused by a high-velocity, non-spherical ``clump" that partially covered the photosphere.  

%Figure \ref{sn2000cx} shows our R-band image obtained with the Bok telescope on October 18, 2000, showing the SN at 101 days after explosion.

%\clearpage

%\begin{figure}
%\plotone{f3.ps}
%\caption{SN 2000cx in NGC 524.  R-band image obtained on Oct. 18, 2000\label{sn2000cx}}
%\end{figure}

%\clearpage

SN 2001C in PGC 19975 was discovered on 2001 January 4.09 UT \citep{2001IAUC.7555....3P}.  The supernova was located at $\alpha = 06^h59^m36^s.10, \delta = +59^\circ31'01''.6$, which is $14''.7$ east and $5''.7$ south of the center of PGC 19975.  It was confirmed to be a Type Ia supernova 13 $\pm$ 2 days after maximum from a spectrum obtained on 2001 January 14.53 UT \citep{2001IAUC.7563....2M}.  There are no other published observations of this SN.  Applying the Lira relation \citep{1999AJ....118.1766P} to our data gives a reddening of $E(B-V) = 0.13 \pm 0.16$.  SN 2001C is categorized as normal based on its spectrum since there is not enough early data to determine the  $\Delta m_{15}(B)$ value.
%Figure \ref{sn2001C} shows SN 2001C 36 days after explosion in a V-band image obtained on Jan 19, 2001 with the Kuiper 61" telescope.

%\clearpage

%\begin{figure}
%\plotone{f4.ps}
%\caption{SN 2001C in PGC 19975.  V-band image obtained on Jan. 19, 2001, 36 days after explosion. \label{sn2001C}}
%\end{figure}

%\clearpage

SN 2001V was discovered in NGC 3987 on 2001 February 19.38 UT \citep{2001IAUC.7585....1J}.  The SN was located at $\alpha = 11^h57^m24^s.93, \delta = +25^\circ12'09''.0$, which is $52''$ east and $28''$ north of the nucleus of NGC 3987.  The spectrum showed a Type Ia supernova well before maximum light. \citet{2003A&A...397..115V} performed early observations of SN 2001V and determined $\Delta m_{15}(B)$ value to be $0.9 \pm 0.1$ and the reddening to be $E(B-V) = 0.05 \pm 0.02$.  

% Figure \ref{sn2001V} shows our March 3, 2001 V-band image, obtained with the Bok telescope, showing the SN at 16 days after explosion. 

%\clearpage

%\begin{figure}
%\plotone{f5.ps}
%\caption{SN 2001V in NGC 3987.  V-band image obtained on March. 3, 2001, 16 days after explosion. \label{sn2001V}}
%\end{figure}

%\clearpage

SN 2001bg in NGC 2608 was discovered on 2001 May 8.943 UT \citep{2001IAUC.7621....1H}.  The supernova was located $22''$ east and $19''$ south of the center of NGC 2608 at $\alpha = 08^h35^m18^s.86, ~\delta = +28^\circ28'05''.8$.  Spectra obtained on 2001 May 10 UT showed the object to be a Type Ia supernova near maximum light \citep{2001IAUC.7622....2G,2001IAUC.7622....3K}.  This SN was observed at the early epochs by KAIT.  The $\Delta m_{15}(B)$ value determined from the KAIT data, provided to us by W-D. Li,  is $0.94 \pm 0.02$, and applying the Lira relation to the early data gives a reddening of $E(B-V) = 0.28 \pm 0.01$. 

% Figure \ref{sn2001bg} shows our January 7, 2002 V-band image, obtained with the Mt. Kuiper telescope, showing the SN at 261 days after explosion.

%\clearpage

%\begin{figure}
%\plotone{f6.ps}
%\caption{SN 2001bg in NGC 2608.  V-band image obtained on Jan. 7, 2002, 261 days after explosion.\label{sn2001bg}}
%\end{figure}

%\clearpage

SN 2001dp was discovered in NGC 3953 on 2001 August 12.9  \citep{2001IAUC.7683....1M}.  The supernova was located at $\alpha = 11^h53^m45^s.1, ~\delta = +52^\circ20'57''.8$, which is $24''$ west and $81''$ north of the center of NGC 3953.  Spectra obtained on August 15.5 UT looked similar to that of Type Ia supernova 1994D one month after maximum \citep{2001IAUC.7683....2A}.  There are no other published observations of this SN.  Our data does not fall in the range needed to apply the Lira relation to determine the host galaxy reddening.  The Galactic reddening in the direction of SN 2001dp is $E(B-V) = 0.03$ \citep{1998ApJ...500..525S}.  SN 2001dp is categorized as normal based on its spectrum since there is not enough early data to determine the  $\Delta m_{15}(B)$ value.

% Figure \ref{sn2001dp} shows our V-band image obtained with the Bok telescope on December 4, 2001, showing the SN at 161 days after explosion. 

%\clearpage

%\begin{figure}
%\plotone{f7.ps}
%\caption{SN 2001dp in NGC 3953.  V-band image obtained on Dec. 4, 2001, 161 days after explosion.\label{sn2001dp}}
%\end{figure}

%\clearpage

\section{Data Analysis}

All data were reduced with the NOAO IRAF software package\footnote{IRAF is distributed by NOAO, http://iraf.noao.edu/ }.  The images were bias-subtracted, flat-fielded using dome flat images, and overscan strip corrected, all done with tasks in the CCDRED package.  The images were then cleaned of cosmic rays using the COSMICRAYS task in the CRUTIL package. 

We performed aperture photometry, in the IRAF package DAOPHOT, on standard star fields from \citet{1992AJ....104..340L} for each night, and the transformation equations were solved using the tasks in the PHOTCAL package. For each supernova field, the images in the same filters were combined and aperture photometry with aperture correction \citep[cf.][]{1989PASP..101..616H} was performed on a set of local stars.  Those stars were used for relative photometry for nights that were determined to be not photometric.  In those cases, the SN instrumental magnitude was linearly shifted an amount equal to the difference in instrumental magnitude and apparent magnitude of the local standard stars.  The error bars for the SNe were determined by combining the photometric error given by IRAF in quadrature with the average deviation from the local standards.  

%The positions and magnitudes of the local standard stars used in each SN field can be found in Tables \ref{2000Elocals}, \ref{2000celocals}, \ref{2000cxlocals}, \ref{2001Clocals}, \ref{2001Vlocals}, \ref{2001bglocals}, \ref{2001dplocals}. 

\begin{deluxetable}{cccccc}
\tablecolumns{6}
\tablewidth{0pc}
\tablecaption{Photometry of Supernovae\label{snmags}}
\tablehead{
\colhead{JD} & \colhead{Epoch\tablenotemark{a}} & \colhead{B} & \colhead{V} & \colhead{R}& \colhead{I}}
\startdata 
\cutinhead{SN 2000E}
2451836.8& 278&  \nodata    & 20.14 (0.05)& 20.64 (0.06)& \nodata      \\
2451865.7& 307& 20.89 (0.07)& 20.55 (0.03)& 20.97 (0.06)& 20.12 (0.05) \\
2452029.0& 470& 23.45 (0.31)& 23.52 (0.22)& 23.59 (0.07)&  \nodata     \\
2452043.9& 485& 23.47 (0.19)& 23.34 (0.18)& 23.48 (0.23)& 22.31 (0.23) \\
2452061.9& 503& 24.11 (0.18)& 24.03 (0.23)& 23.34 (0.24)& 21.84 (0.26)  \\
2452193.7& 635& 25.82 (0.67)& 24.27 (0.54)&   \nodata   & \nodata       \\
\cutinhead{SN 2000ce}
2451866.0& 218& 22.03 (0.15)& 21.65 (0.15)& 22.32 (0.15)& 21.53 (0.10) \\
2451972.7& 325& 23.35 (0.48)& 23.24 (0.15)& 23.40 (0.27)& 23.56 (0.50) \\
2452028.7& 381& 24.01 (0.34)& 23.66 (0.31)& 24.25 (0.30)&  \nodata     \\
2452043.7& 396& 24.12 (0.40)& 23.84 (0.26)& 23.77 (0.25)& 22.84 (0.26) \\
\cutinhead{SN 2000cx}
2451836.8& 101& 17.42 (0.33)& 17.02 (0.14)& 16.91 (0.12)&   \nodata    \\
2451865.7& 130& 17.90 (0.16)& 17.74 (0.10)& 17.88 (0.12)& 18.42 (0.21) \\
2451928.7& 193&   \nodata   & 19.01 (0.10)& 19.43 (0.10)&   \nodata    \\
2452193.8& 485& 22.58 (0.22)& 22.41 (0.12)& 23.71 (0.20)& 22.72 (0.35) \\
2452235.6& 500&  \nodata    & 23.17 (0.58)&   \nodata   &   \nodata  \\
2452253.6& 518& 23.31 (0.64)& 23.24 (0.65)&   \nodata   &  \nodata   \\
\cutinhead{SN 2001C}
2451928.8& 36&  16.04 (0.17)& 15.28 (0.15)& 15.17 (0.07)& 15.37 (0.13) \\
2451971.7& 80&  17.75 (0.19)& 17.05 (0.16)& 16.63 (0.23)& 16.57 (0.18) \\
2451989.7& 98&  18.05 (0.26)& 17.45 (0.20)& 17.23 (0.13)& 17.45 (0.39) \\
2452193.0& 301& 21.00 (0.23)& 20.80 (0.10)& 21.74 (0.07)& 21.33 (0.11)  \\
2452234.9& 343& 21.29 (0.69)& 21.40 (0.26)& 22.03 (0.19)& 21.45 (0.20)  \\
2452246.8& 355&  \nodata    & 21.59 (0.23)& 22.13 (0.15)& 22.25 (0.29)  \\
2452252.9& 361& 21.80 (0.11)& 21.69 (0.11)& 22.29 (0.09)& 21.83 (0.25)  \\
2452295.8& 404& 22.32 (0.14)& 22.24 (0.09)& 23.14 (0.23)& 22.11 (0.21)  \\
2452323.7& 432& 22.64 (0.35)& 22.89 (0.21)& 23.57 (0.46)& 23.23 (0.88)  \\
\cutinhead{SN 2001V}
2451972.7 & 16  & 14.69 (0.19)& 14.50 (0.17)& 14.61 (0.09)& 14.402 (0.40) \\
2451989.7 & 34  & 15.98 (0.21)& 15.28 (0.12)& 15.27 (0.04)& 15.634 (0.20) \\
2452028.7 & 72  & 17.87 (0.07)& 16.86 (0.08)& 16.57 (0.13)& 16.324 (0.21) \\
2452061.9 & 105 & 18.32 (0.16)& 17.74 (0.16)& 16.91 (0.45)& 17.388 (0.13) \\
2452295.8 & 340 & 21.95 (0.19)& 21.70 (0.23)& 22.32 (0.14)& 20.810 (0.18) \\
2452311.8 & 355 & 21.58 (0.17)& 21.80 (0.07)& 22.17 (0.18)& 21.430 (0.18) \\
2452324.8 & 368 & 22.60 (0.41)& 21.96 (0.08)& 23.09 (0.20)& 21.843 (0.10) \\
\cutinhead{SN 2001bg}
2452282.8& 261& 20.13 (0.26)& 20.28 (0.11)& 20.93 (0.08)& 20.48 (0.14) \\
2452296.8& 275& 20.57 (0.11)& 20.35 (0.12)& 20.98 (0.06)& 20.57 (0.12) \\
2452311.8& 290& 20.61 (0.20)& 20.46 (0.26)& 21.23 (0.11)& 20.83 (0.13) \\
2452324.8& 303& 20.67 (0.23)& 20.62 (0.10)& 21.42 (0.16)& 20.63 (0.12) \\
2452366.7& 345&  \nodata    & 21.36 (0.14)& 22.05 (0.11)&  \nodata     \\
2452367.6& 346&  \nodata    & 21.33 (0.27)& 22.30 (0.12)&   \nodata    \\  
2452399.7& 378& 21.99 (0.22)& 22.10 (0.20)& 22.61 (0.17)& 21.61 (0.21) \\
\cutinhead{SN 2001dp}
2452247.8 & 161 &17.78 (0.51) & 17.69 (0.26)& 17.83 (0.07)& 17.75 (0.05) \\
2452282.8 & 195 &18.24 (0.55) & 18.31 (0.05)& 18.62 (0.05)& 18.33 (0.06) \\
2452311.8 & 224 &18.89 (0.25) & 18.83 (0.05)& 19.14 (0.04)& 18.67 (0.04) \\
2452324.8 & 237 &19.09 (0.11) & 19.22 (0.27)& 19.34 (0.04)& 18.78 (0.03) \\
2452366.7 & 279 &  \nodata    & 19.44 (0.23)& 19.99 (0.04)& \nodata      \\ 
2452367.6 & 280 &19.83 (0.23) & 19.73 (0.05)& 20.03 (0.04)&  \nodata     \\
2452399.7 & 312 &20.40 (0.19) & 20.08 (0.07)& 20.50 (0.06)& 19.62 (0.04) \\
2452434.8 & 347 &   \nodata   & 20.62 (0.06)& 21.18 (0.07)&   \nodata   \\ 
2452458.8 & 371 &  \nodata    & 21.49 (0.29)&  \nodata    &   \nodata   \\
2452459.8 & 372 &  \nodata    &  \nodata    & 21.38 (0.30)&    \nodata   \\

\enddata
\tablenotetext{a}{Days since explosion assuming an 18d rise time.}

\end{deluxetable}

\clearpage

\subsection{Image Subtraction}
Some of the supernovae in this project were in very complicated regions of their host galaxies.  If standard aperture photometry was done on these objects, the aperture and sky annulus would be contaminated with galaxy light giving magnitudes for the SNe that are incorrect.  To eliminate this problem, we chose to do template image subtraction, where images of each galaxy are taken after the supernova has faded and are used to subtract the galaxy light from the SN image.  For each of these supernovae, except SN 2000cx, image subtraction was performed.  SN 2000cx was located far from the nucleus of an S0 galaxy, so it was determined that the galaxy light could be acceptably minimized without subtraction.  For each SN, we employed aperture photometry with aperture correction. 

\begin{figure}
\plottwo{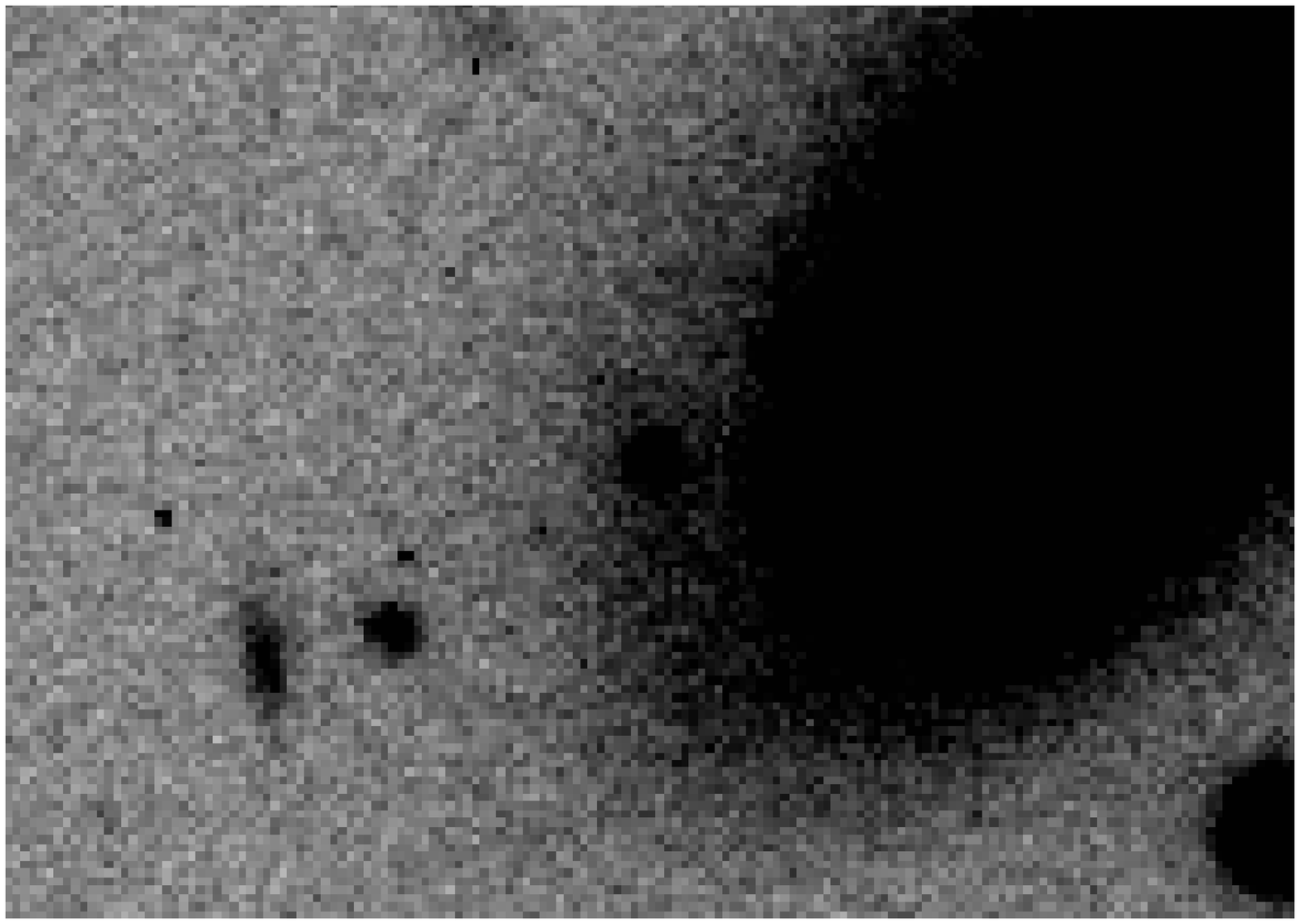}{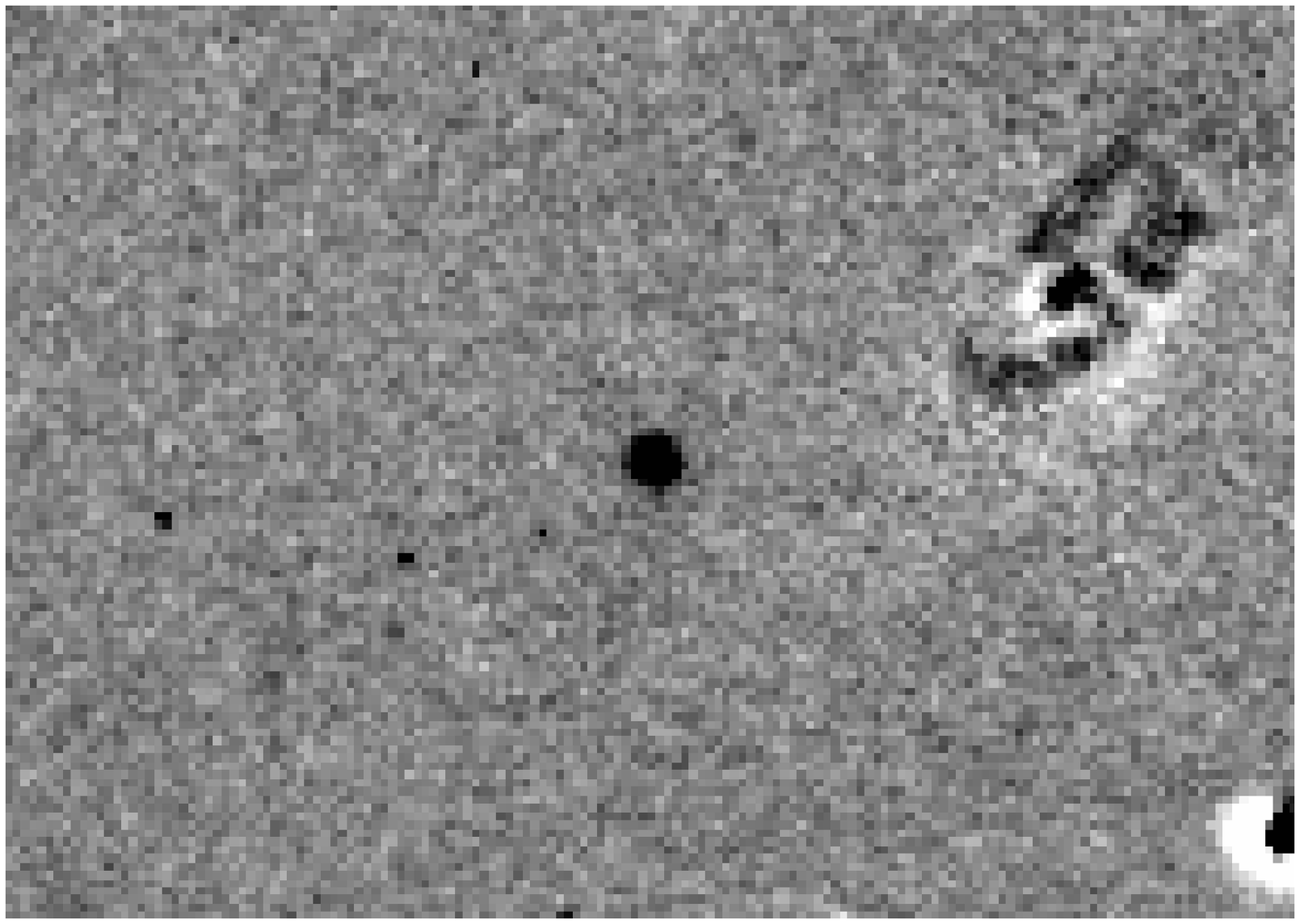}
\caption{SN 2001C on Oct 10, 2001, 301 days after explosion, before (left) and after (right) subtraction.  The SN is in the center of both images.  \label{sn2001Csub}}
\end{figure}

There are several steps to prepare the images for subtraction.  First, the images were aligned with the tasks in the IMGEOM package.  Then, if necessary, the images were convolved with a Gaussian to match the seeing conditions. Then, if the SN image and subtraction image were obtained with different instruments, the subtraction image is stretched using the MAGNIFY task to match the plate scales.  The images were then scaled to match in intensity and subtracted using IMARITH\footnote{A parallel analysis was performed using the ISIS software \citep{2000A&AS..144..363A} to generate subtracted images.  The photometry from that analysis agrees with the photometry presented in this paper.  The ISIS photometry will be presented in the dissertation of J. Lair.}.  Figure \ref{sn2001Csub} shows an example of image subtraction.  The image on the left is SN 2001C, obtained when the SN was 301 days after explosion.  The SN is in the center of the image surrounded by galaxy light.  The right panel shows the image after subtraction.  Again the SN is in the center of the image, but now the surrounding galaxy light has been subtracted out.  The subtraction is clearly not a perfect process.  The very center of the galaxy and some bright point sources do not subtract out completely, but the diffuse galaxy light that would contaminate the aperture and sky annulus is removed.

Photometry was performed on the supernova in the subtracted image using a small aperture with aperture correction.  The aperture correction done using the MKAPFILE task.  For SN 2000cx, the photometry was done with aperture correction without subtraction.  Table \ref{snmags} shows our measured magnitudes for the SNe in this study. 

\clearpage

\clearpage

\section{Results}

\begin{figure}
\plotone{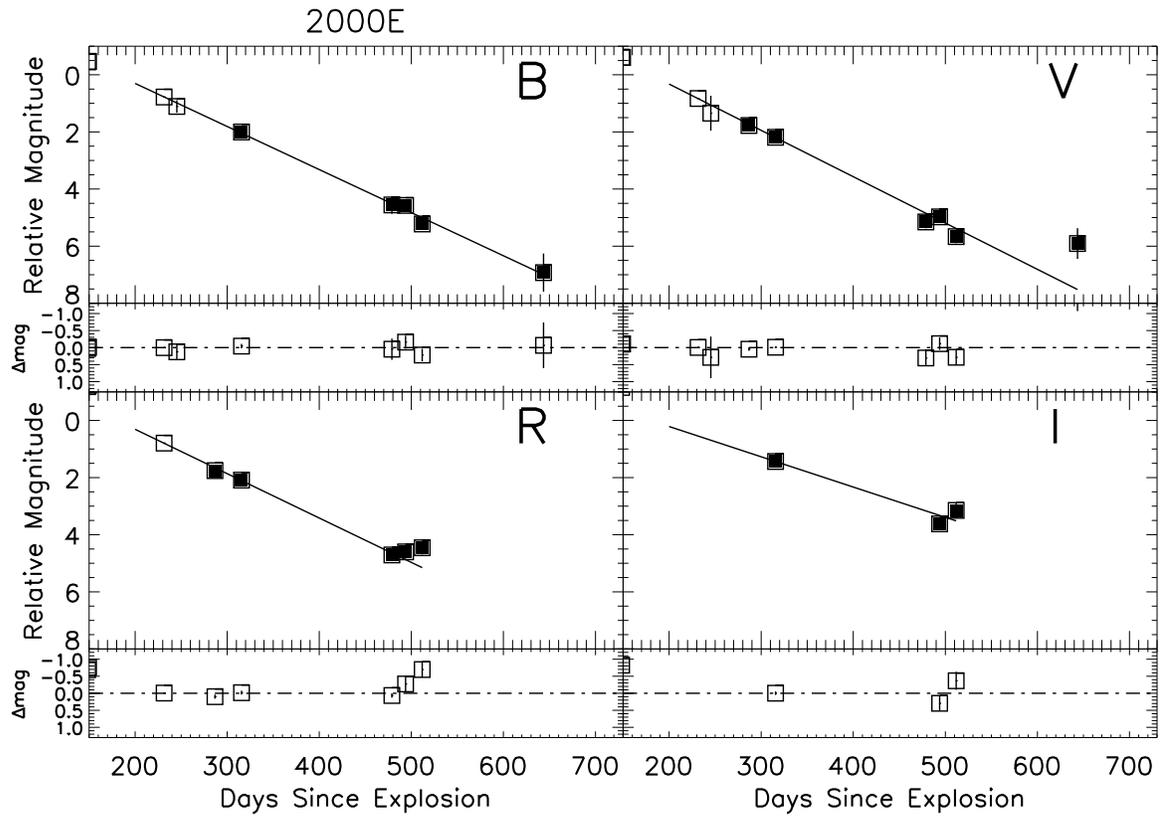}
\caption{SN 2000E light curves.  The open symbols are data from \citet{2003ApJ...595..779V}\label{sn2000E_slope}}
\end{figure}

\begin{figure}
\plotone{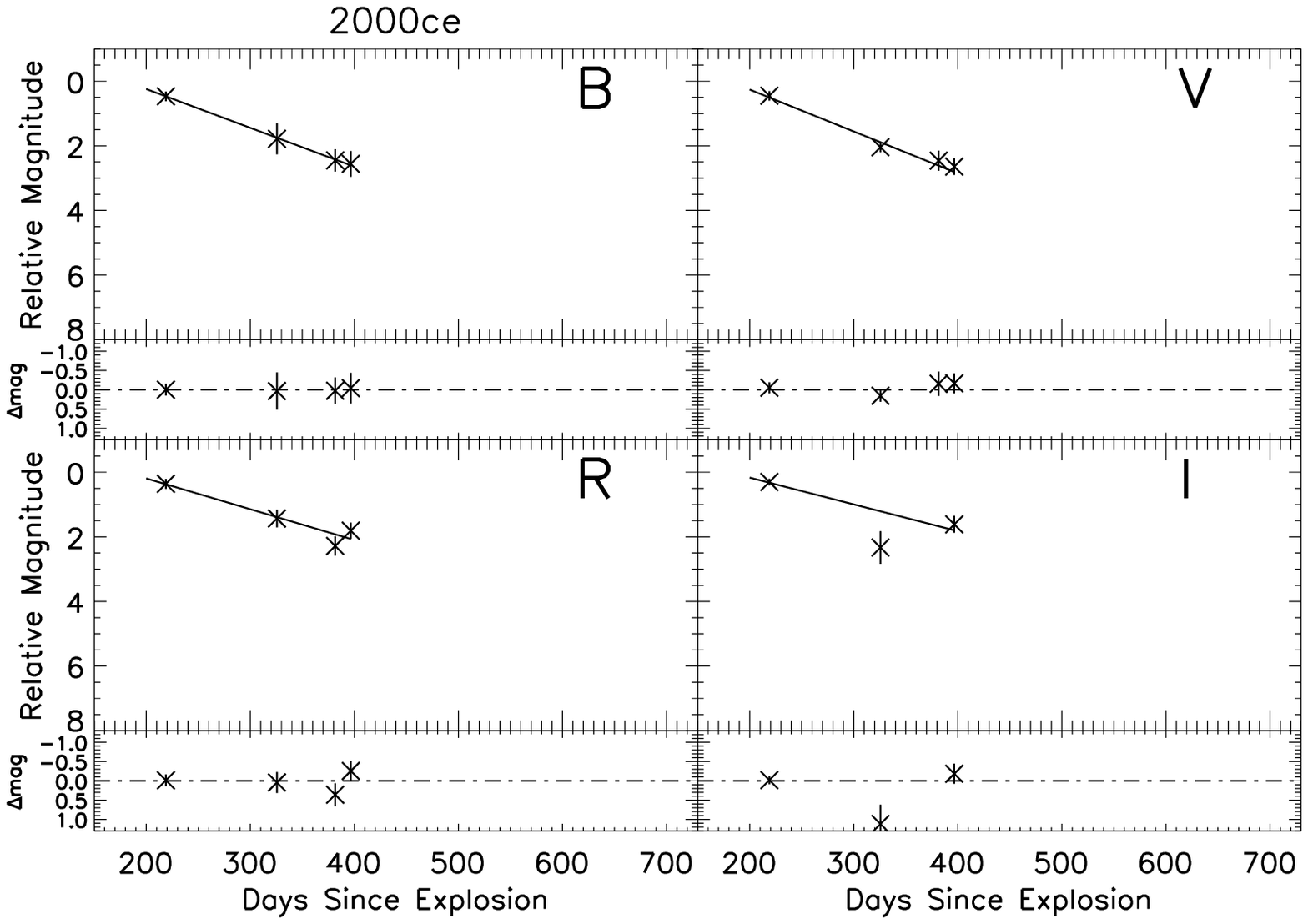}
\caption{SN 2000ce light curves.\label{sn2000ce_slope}}
\end{figure}

\begin{figure}
\plotone{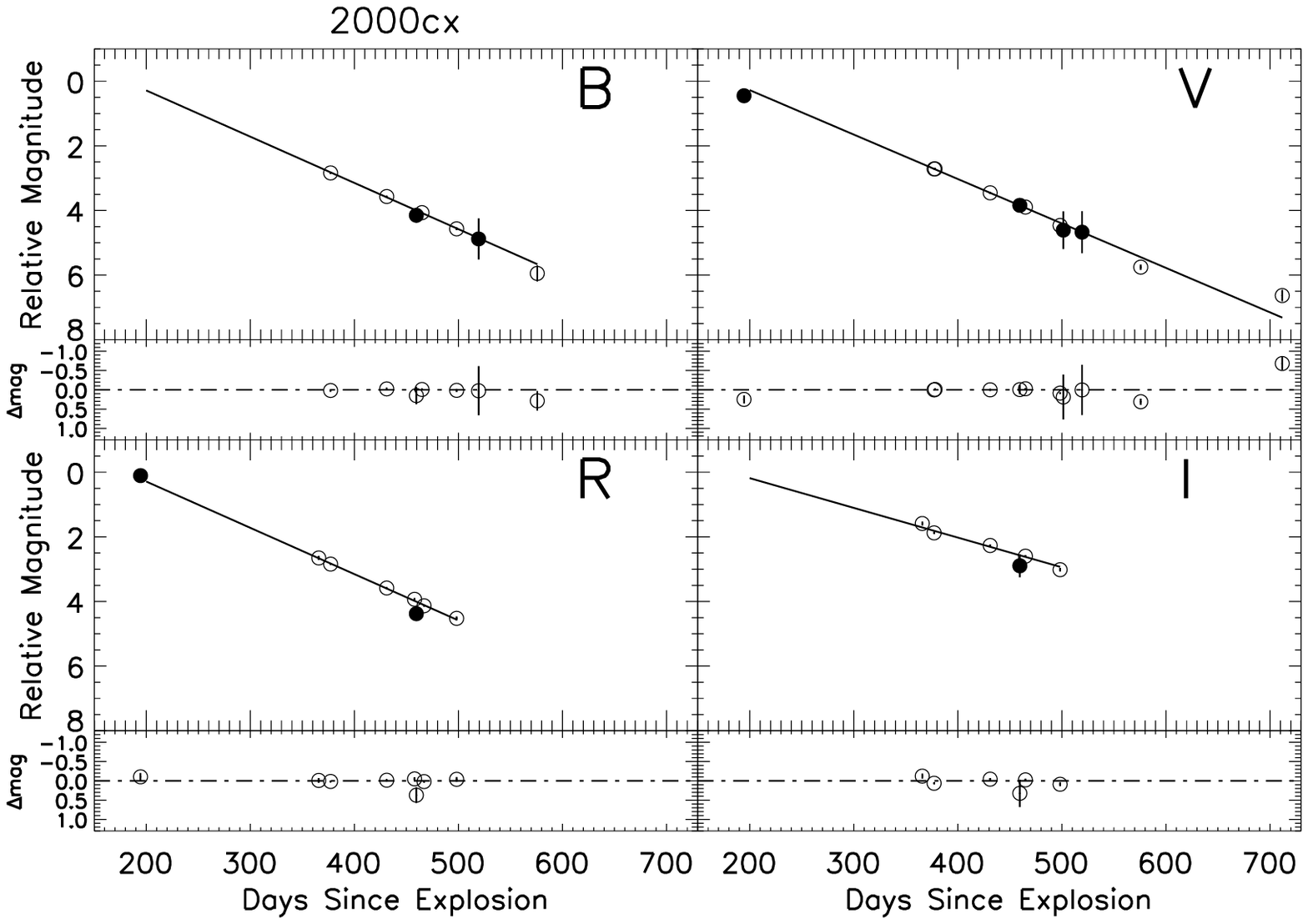}
\caption{SN 2000cx light curves.\label{sn2000cx_slope}}
\end{figure}

\begin{figure}
\plotone{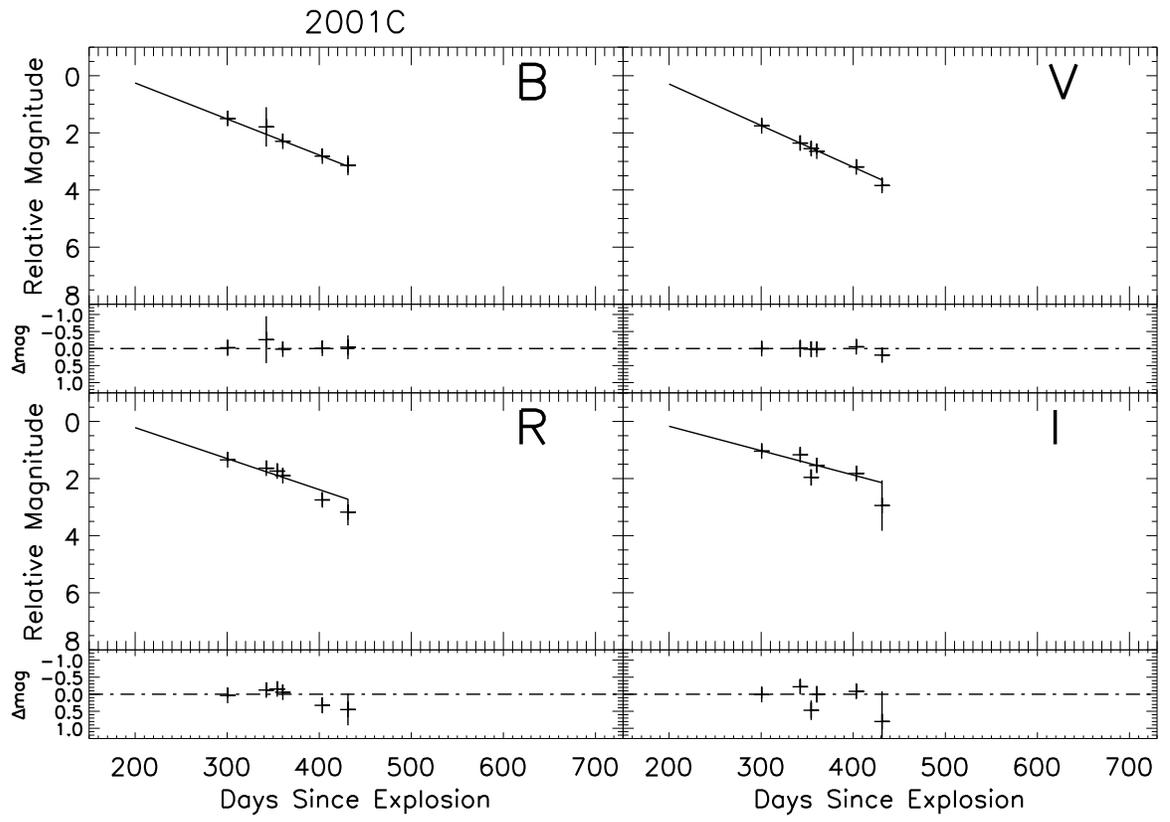}
\caption{SN 2001C light curves.\label{sn2001C_slope}}
\end{figure}

\begin{figure}
\plotone{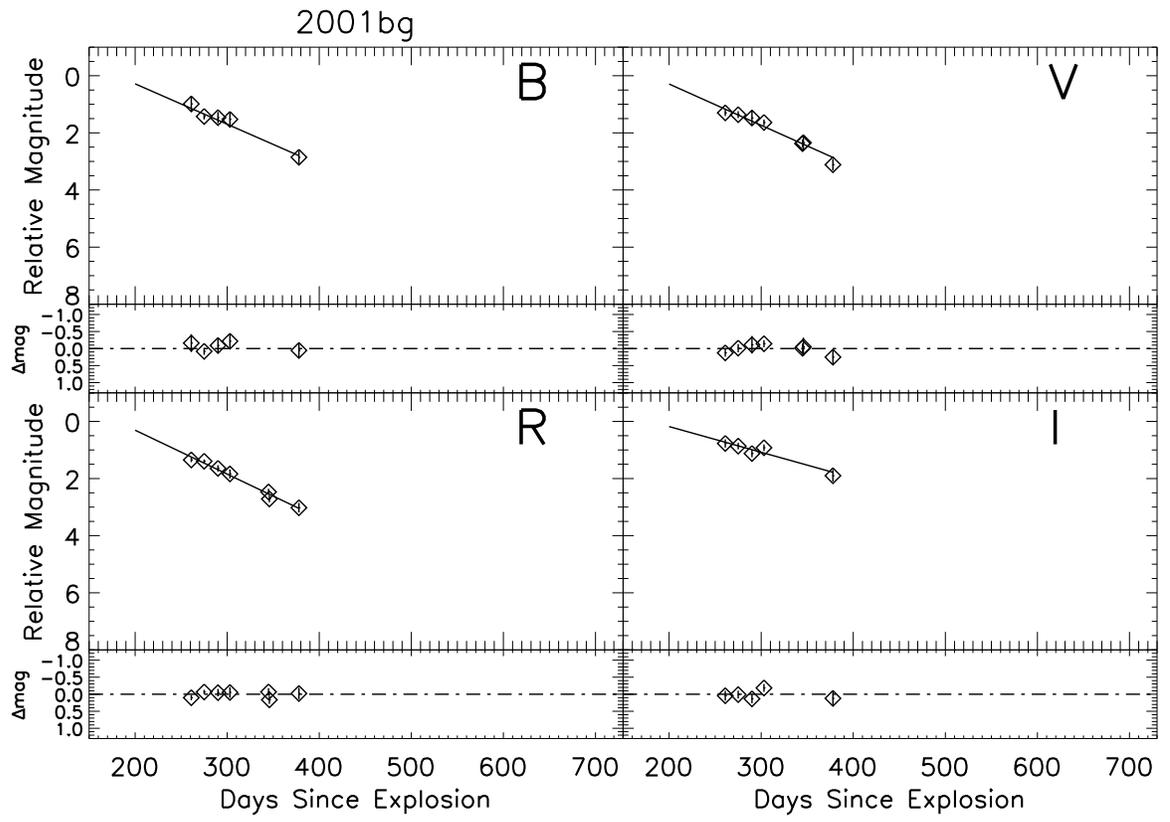}
\caption{SN 2001bg light curves.\label{sn2001bg_slope}}
\end{figure}

\begin{figure}
\plotone{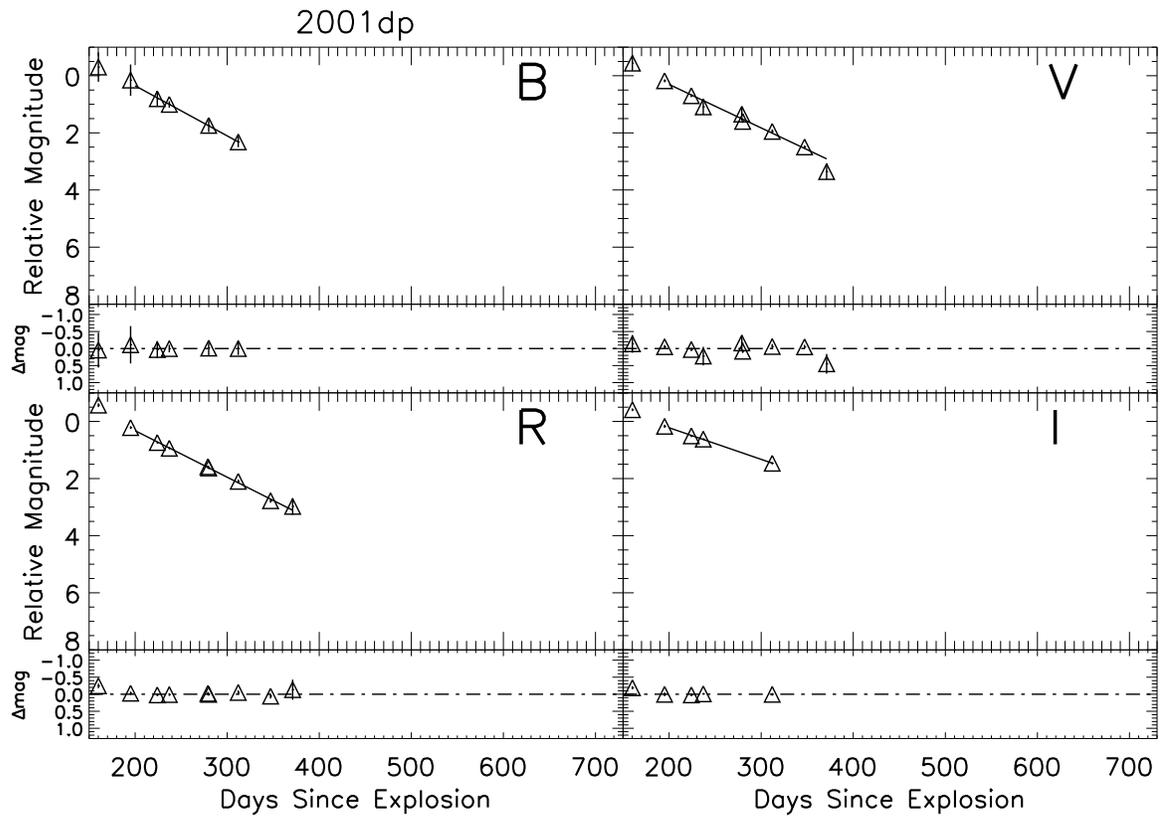}
\caption{SN 2001dp light curves.\label{sn2001dp_slope}}
\end{figure}

\begin{figure}
\plotone{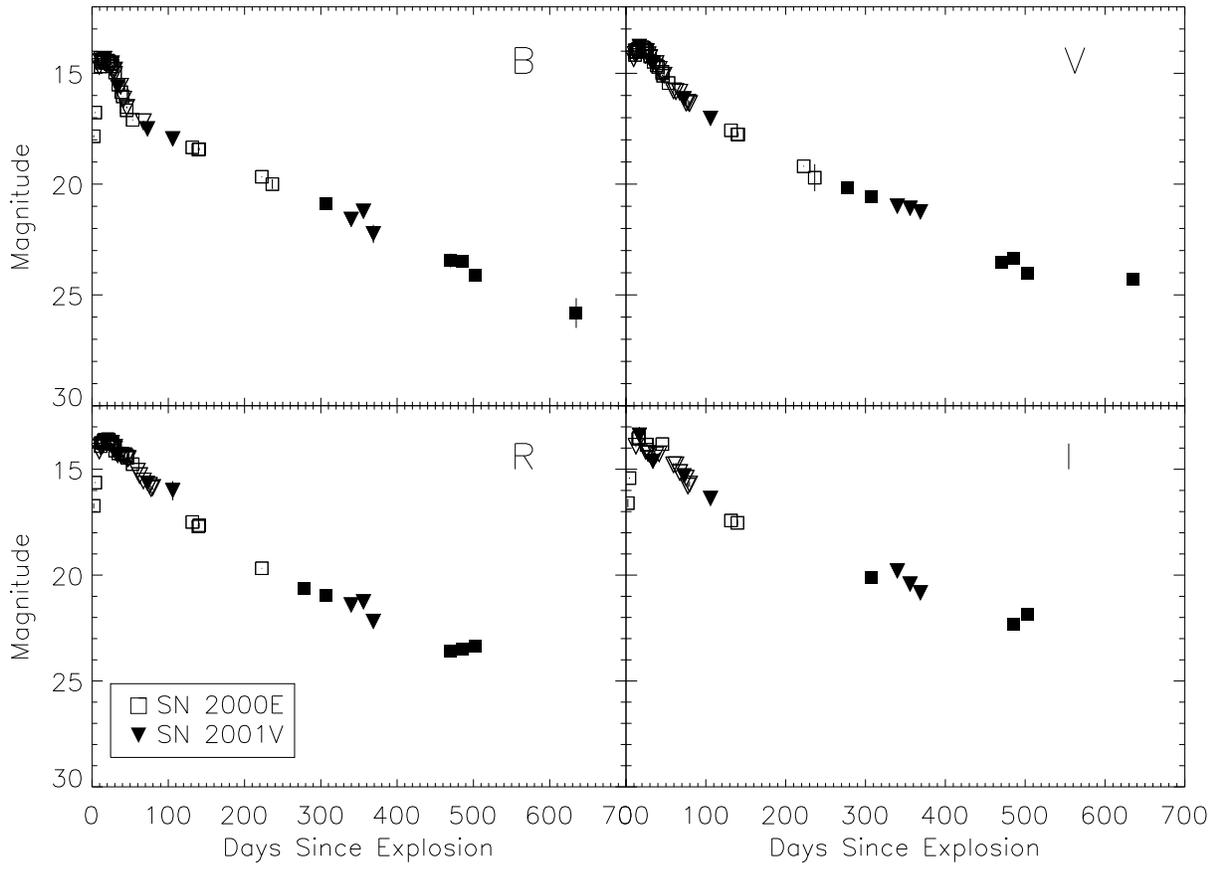}
\caption{SN 2001V light curve normalized to the peak of SN 2000E.\label{sn2001V_comp}}
\end{figure}

The light curves of the SN 2000E, SN 2000ce, SN 2000cx, SN 2001C, SN 2001bg, SN 2001dp can be seen in Figures 
\ref{sn2000E_slope}, \ref{sn2000ce_slope}, \ref{sn2000cx_slope}, \ref{sn2001C_slope}, \ref{sn2001bg_slope}, 
\ref{sn2001dp_slope}, respectively.  The data are normalized to be zero magnitude at 200d past explosion assuming an 18 day rise time to peak magnitude.  The solid line in each figure is a linear fit to the data between 200 and 500 days. The figures also show a residual plot showing the deviation of the SN magnitude from the value of the fitted line. In Figure \ref{sn2000E_slope}, the open symbols are data from \citet{2003ApJ...595..779V} and the closed symbols are our data.  In Figure \ref{sn2000cx_slope}, the open symbols are a combined data set from \citet{2001PASP..113.1178L}, \citet{2004A&A...428..555S}, and \citep{2002PhDT........10J}.  The SN 2001V data set did not include a sufficient amount of data in the 100-400d range to allow for a linear fit to the data.  
Figure \ref{sn2001V_comp} shows the light curves of SN 2001V plotted with the light curves of SN 2000E.  The SN 2001V data was normalized to the peak of SN 2000E.  The early time data of SN 2001V, plotted with the open symbols, is from \cite{2003A&A...397..115V}.

\begin{figure}
\plotone{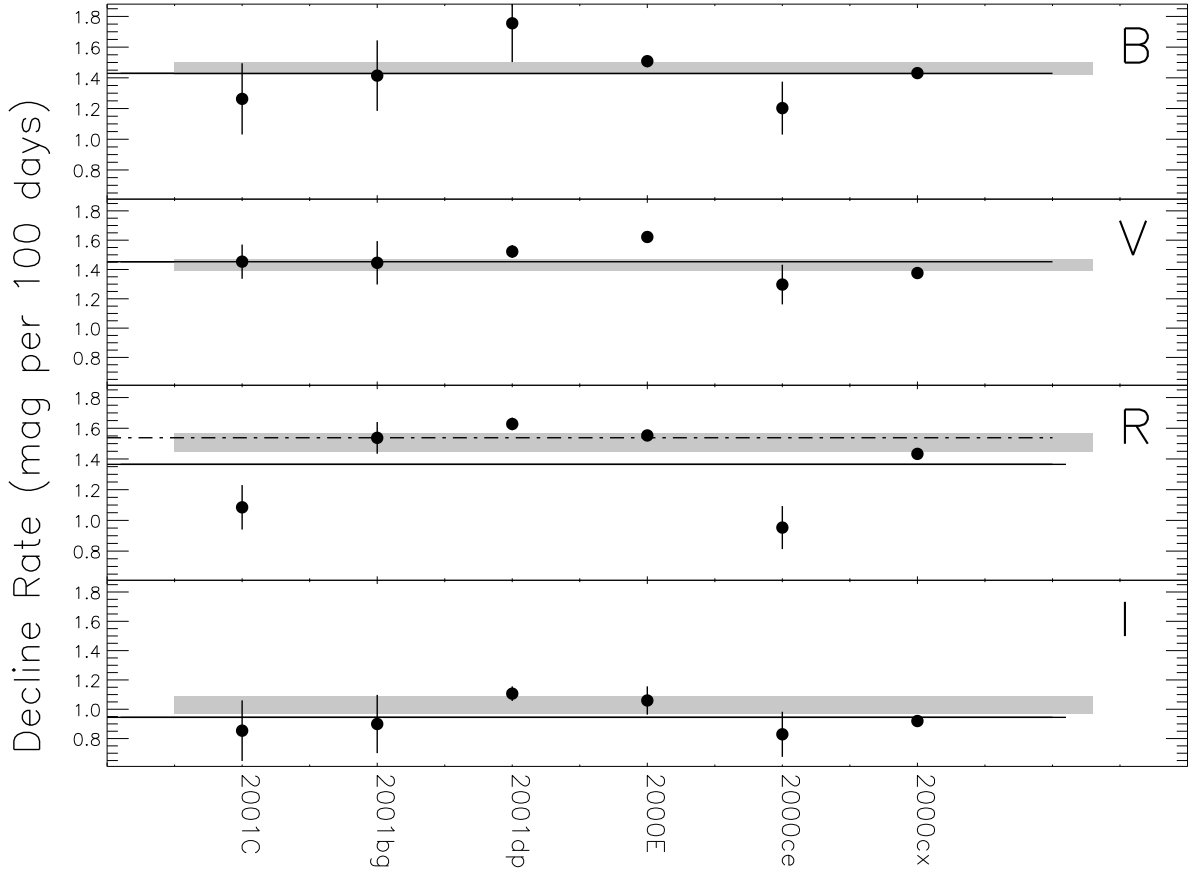}
\caption{\label{aveslope}Average slopes of SN light curves fit between 200-500 days.  The solid line is the average of the SNe in this study and the shaded bar is the average slope with  1$\sigma$ error of 12 N/SP SNe from MTL01.  The dash-dot line is a re-calculation of the average leaving out SN 2000ce and SN 2001C.}
\end{figure}

From these plots we see no systematic trend that would suggest anything other than a linear decline during those epochs.  Figure \ref{aveslope} shows the slopes of the linear fit to the data for each SN along with the calculated 
average of all the SNe.  The shaded bar is the average slope of 16 N/SP SNe Ia from MTL01 with a 1$\sigma$ error.  The dash-dotted line is the average R-band slope having removed SN 2000ce and SN 2001C.  This was done just to show there is a consistency with the MTL01 slopes without those two objects.  The calculated averages can be seen Table \ref{decline} along with the decline rates of the models during this epoch (for comparison, the $^{56}$Co decay rate gives 0.96 mag/100d)  The SN decline rates, the BVR light curves being around 1.4 mag per 100 days and the I-band being much shallower at 0.94 mag per 100 days, are consisted with the results of \citet{2004A&A...428..555S} for SN 2000cx.  This shows that the SN 2000cx light curves are not anomalous at late times and the shallower slope in the I-band that was seen in SN 2000cx is a general feature of the light curves of N/SP SNe Ia.  The slower decline rate of the R-band of SN 2000ce and SN 2001C cannot be fully explained without late time spectra of those objects.  The R-band contains 17\% of the BVRI flux in the SN 2000ce and SN 2001C light curves, as opposed to 11\% in SN 2000E and SN 2000cx.  It is clear, for example, V-band light curves cannot be used alone to trace the bolometric evolution for all SNe.  Still, summing the BVRI fluxes gives a luminosity steeper than the $^{56}$Co decay due to the deposition fraction of the gamma-rays decreasing during the epoch.   

\clearpage

\clearpage

\clearpage

\clearpage

\clearpage

\clearpage

\clearpage

\clearpage

\clearpage

\begin{deluxetable}{cccc}
\tablecolumns{4}
\tablewidth{0pc}
\tablecaption{Decline Rates (magnitudes per 100d).\label{decline}}
\tablehead{
\colhead{B} & \colhead{V} & \colhead{R} & \colhead{I}  }
\startdata 
\cutinhead{Average of SNe in this study}
1.43 (0.07)& 1.46 (0.04) & 1.36(0.04)& 0.94 (0.06) \\
\cutinhead{Kozma simulations with photoionization}  
1.19 & 1.29 & 1.43 & 1.42 \\ 
\cutinhead{Kozma simulations without photoionization}
%\multicolumn{4}{c}{Kozma simulations without photoionization} \\ \hline
1.75 & 1.73 & 1.47 & 1.25 \\
\cutinhead{MTL01 energy deposition}  
\multicolumn{2}{l}{Positron escape}  1.31 \\
\multicolumn{2}{l}{Positron trapping}  1.11 \\ 
\enddata

\end{deluxetable}

\clearpage

\section{Discussion}

\begin{figure}
\plotone{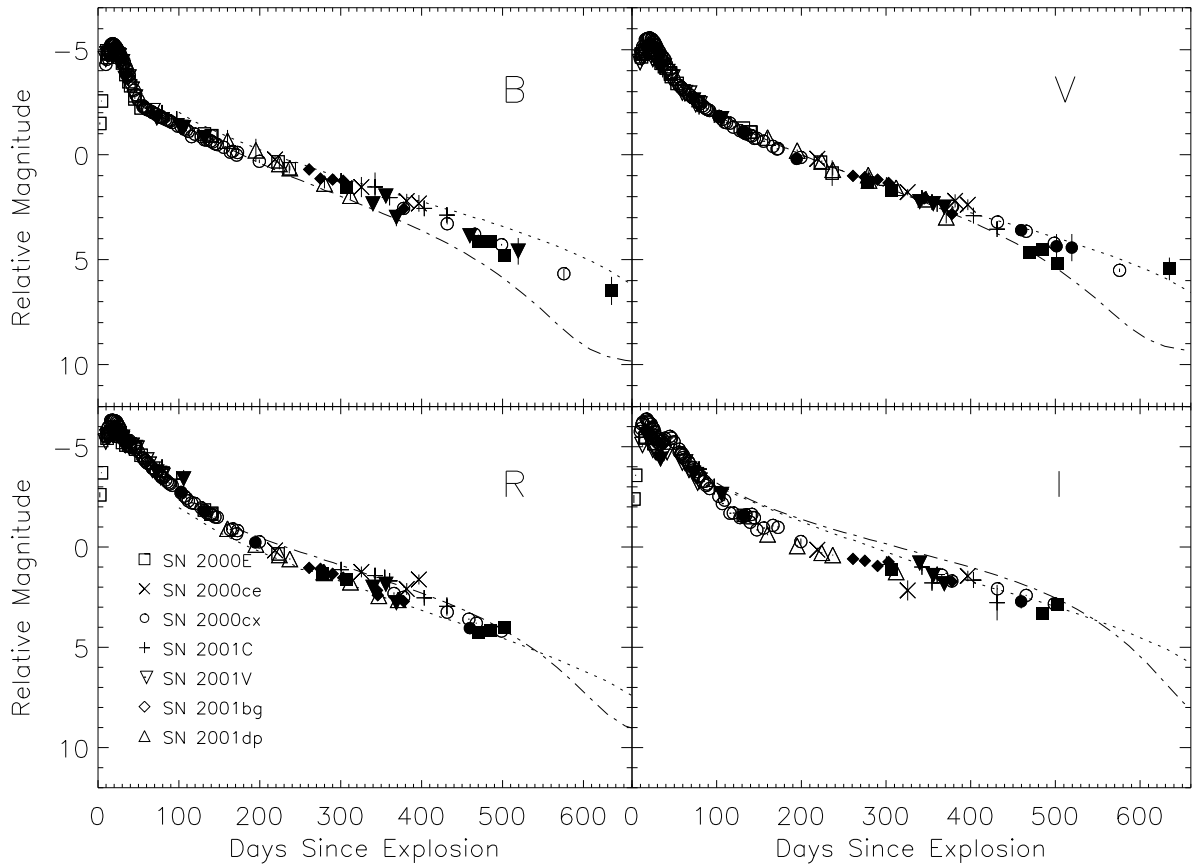}
\caption{Radiation transport models of \citet{2004A&A...428..555S}.  The model including photoionization is plotted as the dotted curve and the model with photoionization is plotted with the dot-dashed curve.\label{radmodel}}
\end{figure}

The model light curves of \citet{2004A&A...428..555S}, where they performed radiation transport on the SN Ia model W7 \citep{1984ApJ...286..644N} with complete and instantaneous positron trapping are shown in Figure \ref{radmodel} plotted with the SN data [on the model curves we have also included the early time KAIT data of SN 2001bg provided by W-D. Li \citetext{priv. comm.}].  Sollerman et al. fit these models to SN 2000cx only, so the models were corrected to the distance and extinction of that object.  We removed those corrections from the models and from that normalized the V-band model curves to be zero magnitude at 200d past explosion along with the data.  The B,R, \& I band model curves were then adjusted to preserve the colors of the model calculation; the colors of the models at 200d were set to be equal to the colors of SN 2000cx at 200d.   The dotted curve is their model including photoionization and the dash-dotted curve is the model without photoionization, representing the two extreme treatments of UV photons, either ignoring the effects of UV scattering or assuming that all UV photons are redistributed to longer wavelengths.  The B~\&~V band data seem to favor the model including photoionization.  The I-band data is more shallow than both of the model assumptions. The slopes of each simulated light curve are shown in the lower portion of Table \ref{decline}.  For comparison, the slopes of the energy deposition rate simulations of MTL01 are also shown in the lower portion of Table \ref{decline}. As reported previously, the slope of the V and B band light curves do roughly match the positron escape energy deposition rates, but the I band slope is shallower. 

\clearpage

\clearpage

%The data plotted on the positron transport models of MTL01 can be seen in Figure \ref{posmodel}.  The solid line is the energy deposition model with a tangled magnetic field leading to positron trapping and the dashed line is the weak and/or radially combed field model that can allow positron escape.  The data and models are normalized to be zero magnitude at 200d past explosion. In the BVR bands the data fall below the trapping curve, following along the radial curve.  In the I-band, the data is more shallow than both the trapping and radial curves.    

%\clearpage

%\begin{figure}
%\plotone{f18.ps}
%\caption{Positron transport models of MTL99, MTL 01.  the solid curve is the model with a tangled magnetic field leading to positron trapping, and the dashed line is the weak and/or radially combed magnetic field which can allow positron escape.\label{posmodel}}
%\end{figure} 

%\clearpage
The light curves shown in Figure \ref{radmodel} represent data from seven SNe Ia, but show very little deviation from a single curve for each photometric band. This suggests that positron transport and the resulting levels of ionization from these sub-classes of SNe Ia do not occur on a SN-by-SN basis, rather it is more likely that there is a single answer to the question of positron escape. The flatness of the I band data is an optical wavelength confirmation of the fact that color evolution does occur in the light curves of the N/SP SNe Ia and must be accounted for to accurately produce bolometric light curves to compare with energy deposition rates. 

Although the radiation transport simulations provide an acceptable fit to the light curves of the B,V and R bands, the color evolution as simulated does not account for the flatness of the I band light curve.  Efforts to fit the (sparse) sampling of late, nebular spectra of N/SP SNe Ia have demonstrated that the emission after 200 days is dominated by forbidden lines from Fe II and Fe III, as well as from contributions from stable Co II/Co III and Ni II/Ni III. The NIR spectral range contains only lines from singly-ionized iron-peak elements, so the flat NIR light curves would presumably be due to an increase in the fraction of singly-ionized iron-peak elements. If this is indeed the correct explanation for the NIR light curve slopes, it is not clear why the I-band emission follows a slope intermediate to the B,V,R and J,H,K slopes. 

%One explanation could be emission from non iron-peak elements such as calcium, which has a permitted line complex at $\sim$7200$\AA$. However, as the R band is also sensitive to this line feature, it seems impossible to simultaneously explain the shallow I band slope while preserving a steep R band slope. As the slope of the R band light curves for SNe 2001C and 2000ce {\it do} feature shallow slopes roughly equal to the I band slopes, for those two SNe, perhaps a dominant $\sim$7200$\AA$ feature dominates the late R \& I band light curves of these two SNe. Of course, that presents problems for the explanation of the R and I band light curves for the other five SNe Ia, as the I band slopes of SNe 2001C and 2000ce are not anomalous. This topic requires further study, and remains a puzzle for the present. As the nebular spectra of sub-luminous SNe Ia feature a bright emission feature at $\sim$7200$\AA$ relative to the complex of lines from 4000$\AA$~-~5700$\AA$, studying the late B,V,R and I band light curves of that sub-class of SNe Ia should be particularly interesting.   

\section{Summary}
We have presented BVRI photometry of seven normally-luminous Type Ia supernovae at epochs out to $\sim600$ days past explosion, roughly doubling the available data for the late epochs.  The resulting B,V and I band light curves do not show any evidence of inhomogeneity within the N/SP subclasses of SNe Ia.  There is also no evidence that the very late time B \& V band light curves start to rapidly dim, which would indicate the start of an IRC.  However, it is clear from the results of these observations and the NIR observations of SNe 1998bu, 2000cx and 2003cg that more optical/NIR observations of these objects at late times are needed to determine if the shallower I-band slope is due to the emission shifting into the longer wavelengths and to reproduce the flat NIR light curves of SN 2000cx.   

The core question as to whether positrons escape the SN ejecta cannot be answered based only upon the optical photometry of SNe Ia. As the number of SNe Ia detected in the NIR increases, so does the evidence for a NIR excess that is suggestive of positron trapping. This optical study supports previous claims that there is little variation among the light curves of N/SP SNe Ia late times. This lack of variation suggests that if trapping is the better explanation for one N/SP SN Ia, it is the better explanation for all N/SP SNe Ia (at least until which time an anomalous late light curve is detected). Radiation transport simulations have never been performed upon SN models that allow for positron diffusion throughout the ejecta with a fraction escaping the ejecta. Until such simulations are performed and compared to the observations, it is premature to conclude that positrons are trapped inside the SN ejecta. However, it does appear that radiation transport simulations that include models that trap positrons can reproduce the general features of the late SN Ia light curves. It is a challenge for simulators to determine whether the shortcomings of the current simulations point to small issues in the simulations within the positron trapping scenario, or to the need for some degree of positron diffusion and escape.

\end{document}